# Solar radio bursts observations by Egypt- Alexandria CALLISTO spectrometer: First results

F. N. Minta[a,*], S. I. Nozawa[b], K. Kozarev[c], A. Elsaid[d,e], A. Mahrous[a]

[a] *Department of Space Environment, Institute of Basic and Applied Sciences, Egypt-Japan University of Science and Technology, 21934, Alexandria, Egypt.*
[b] *Department of Science, Ibaraki University, 2-1-1 Bunkyo, Mito, Ibaraki 310-8512, Japan.*
[c] *Institute of Astronomy with National Astronomical Observatory, Bulgarian Academy of Sciences, Sofia, Bulgaria.*
[d] *Department of Applied and Computational Mathematics, Institute of Basic and Applied Sciences, Egypt-Japan University of Science and Technology, 21934, Alexandria, Egypt.*
[e] *Mathematics & Engineering Physics Department, Faculty of Engineering, Mansoura University, PO 35516, Mansoura, Egypt.*

Abstract

The newly installed CALLISTO spectrometer, hosted by the Department of Space Environment, Institute of Basic and Applied Sciences- EJUST, commenced operation on August 14, 2021. The system contains a cross dipole long-wavelength array antenna with high sensitivity to monitor solar radio transients. Its antenna was strategically positioned and appeared to be in the center of the CALLISTO network of spectrometers. Moreover, in the northern section of Africa, the Egypt-Alexandria CALLISTO and ALGERIA-CRAAG stations are the only ones operating. There are no stations in the West African region, while stations in the eastern part of Africa are not working. Thus, Egypt- Alexandria station serves as a reference for other stations within the e-CALLISTO network. Despite the low solar activity, the instrument detected several solar radio bursts not limited to type II, type III, and type V. A vigorous case study was conducted on two selected radio burst events to validate the authenticity of the recorded events. Other solar radio stations at different geographical locations recorded all the radio bursts detected by the spectrometer. The case study included brief analyses that indicated a type II radio burst observed on October 09, 2021, between 06:30 and 07:00 UT, was associated with an M1.6 solar flare located at N18E08 within NOAA-AR 12882 and a CME with a shock front speed of ~978 km/s. However, the type III radio burst is neither CME nor solar flare associated. These analyses examine the instrument's capacity to provide real-time solar radio transient data 24 hours a day to mitigate the challenges of data gaps faced in the African continent. Hence, the instrument has become an integral part of space weather monitoring and forecasting over the region and other parts of the globe.

---

[*] Corresponding author.
Felix N. Minta, Email: felix.minta@ejust.edu.eg

## 1.0 Introduction

Coronal mass ejections (CMEs) and solar flares are large-scale energetic activities observed from gamma rays in hundreds of mega electron volts to radio wavelength in tens of meters across the electromagnetic spectrum (Zucca et al., 2012). These events can trigger solar radio transients at metric and decametric wavelengths. Solar radio transients are categorized based on the frequency, wavelengths, and spectral features as; i) Type I, which are short-lived narrow band noise storms associated with active regions and occur within a frequency range of 80 MHz- 200 MHz (Ramesh et al., 2013; Mercier et al., 2015); ii) Type II which generally have a fundamental frequency between 20-150 MHz and are thought to be associated with coronal shocks (Nelson and Melrose 1985; Kumari et al., 2017); iii) Type III are known to be related to solar flares and fast-drifting radio burst within 10 kHz- 1 GHz frequency range (Ratcliffe et al., 2014); iv) Type IV, shows broad fine spectrum within 20 MHz- 2 GHz frequency band and are related to CME and flux loops (Sasikumar Raja et al., 2014; Morosan et al., 2020); v) Type V, short-duration spectra with frequency 10 MHz-200MHz and usually accompanies a type III.

In recent times, unlike interferometers, spectrometers have sufficient frequency channels with high-frequency resolutions to identify various solar radio emissions. Several radio spectrometers have been designed and deployed to monitor and observe solar radio transients. For instance, the Culgoora Radioheliograph was designed to produce radio spectra images within the 18-1800 MHz frequency range (Sheridan et al., 1972). The Nançay Decameter Array (NDA) was designed to produce radio dynamic spectra at 10-80 MHz in conjunction with an operational radioheliograph producing solar radio images between 150- 420 MHz (Dumas et al., 1982; Benz et al., 1991). Others include the Phoenix II, which is based on Fourier Transforms operating between 0.1-4 GHz, the Green Bank Solar Radio Burst Spectrometer (GBSRBS) operating at 18-1100 MHz frequency window, and the Artemis spectrograph observing solar transients at 20- 645 MHz, with a stationary antenna of frequency, 20-110 MHz, and a moving parabolic antenna at 110- 650 MHz (Kontogeorgos et al., 2007; White, 2007).

The recent milestone in solar radio transients monitoring is achieved by deploying a portable and commercially available electronic-based radio telescope known as the Compound Astronomical Low-cost Low-frequency for Spectroscopy and Transportable Observatory (CALLISTO). The frequency-agile spectrometer is designed in a frequency range of 10- 870 MHz to monitor metric and decametric radio transient emissions 24×7 with the ability to explore the solar corona in a heliospheric distance of 1- 3 $R_s$ (Benz et al., 2005,2009; Pohjolainen et al., 2007; Zucca et al., 2012).

Over 56 out of 184 stations within the e-CALLISTO network provide real-time data (15 minutes/ frame) to the central server stationed at the University of Applied Sciences (FHNB) in Bragg/Windisch, Switzerland. The data from the CALLISTO radio spectrometer are in FITS file format and can be transferred using an RS-232 cable to a computer. The data is freely available at http://www.e-callisto.org/Data/data.html according to the open data management policy by NASA and UN International Space Weather Initiative program.

It is noteworthy that a GPS clock regulates the timely operation of CALLISTO in UTC. Consequently, the relative timing is accurate to less than one millisecond, while the absolute timing is unclear to a few milliseconds due to internal delays (Benz et al., 2005 and references therein). Individual channels of the instrument have a radiometric bandwidth of 300 kHz and can be tuned by control software in steps of 62.5 kHz to obtain high spectral resolution.

This article presents a detailed report on the specifications and configuration of the CALLISTO spectrometer in Alexandria- Egypt (section 2), a case study, and analyses of selected radio bursts (section 3) to examine the authenticity of the detected radio bursts and the instrument's capacity to provide real-time data in mitigating the data challenges faced in the African continent. Summary, conclusion, and future works are presented in section 4.

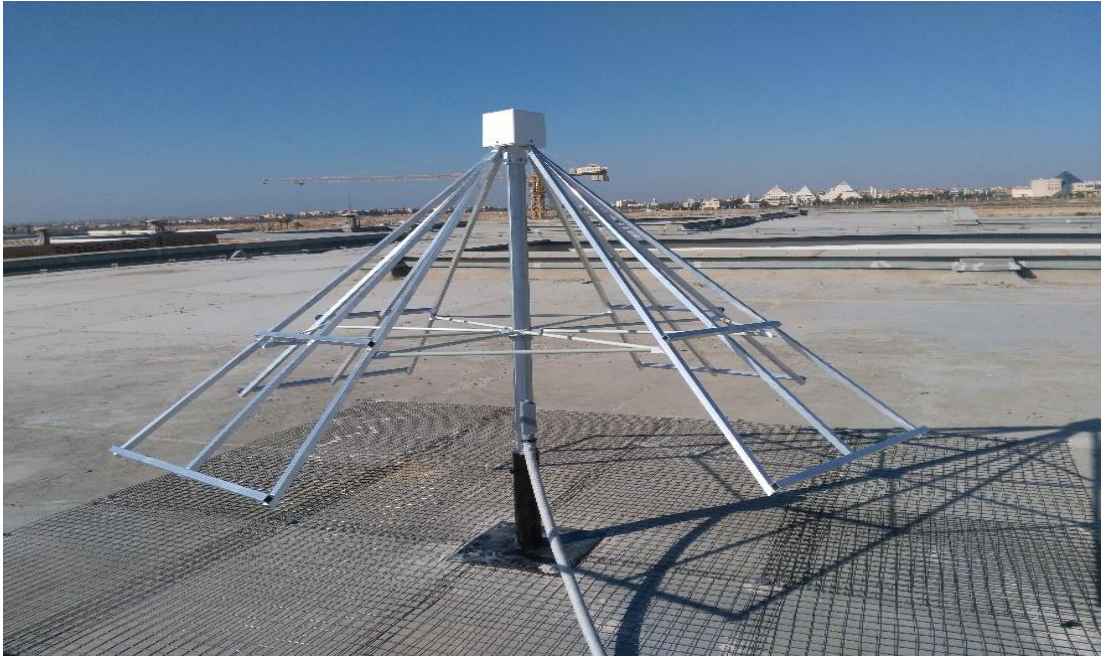

Fig. 1: Installed long-wavelength array antenna on the top of the Space Environment Research Lab-EJUST building at N30⁰, E29⁰. The antenna is an aluminum frame with a CDA, active balun, and ground screen made of a steel grid with a cell size of 2.47 mm x 2.47 mm.

## 2.0 CALLISTO specifications

The CALLISTO system deployed in Alexandria, Egypt, is hosted by the Institute of Basic and Applied Sciences of Egypt-Japan University of Science and Technology. The antenna is installed at a strategic geographical location (N030⁰: 51.69, E029⁰: 33.67) within the middle east to address the radio transient data gaps in developing countries, especially over the African continent. The CALLISTO system comprises a long-wavelength array antenna (LWA), two frequency receivers, dual-channel heterodynes up-converter, power coupler, and a standard windows-based computer. The former, also known as a cross-dipole (CD- LWA) antenna, can withstand severe weather conditions (e.g., 80 m/hr and 100 m/hr gusty conditions), and its maintenance cost is relatively minimal. The other parts of the antenna include the active balun/pre-amplifier at the apex to enhance the system's sensitivity, a support structure, and a ground screen made of steel grid cells of size 2.47 mm × 2.47 mm that prevents back reflection and reduces the susceptibility to changeable soil conditions. The antenna's base length is ~0.5m, and the vertical height is ~1.5m.

Additionally, the LWA antenna consists of two different polarization outputs (East-West and North-South), making it possible to observe the polarization properties of solar flares (Kallunki et al., 2018). The antenna at the final installation stage is shown in Fig. 1. A detailed overview of the components and geometry of the LWA antenna was presented by Hicks et al., (2012).

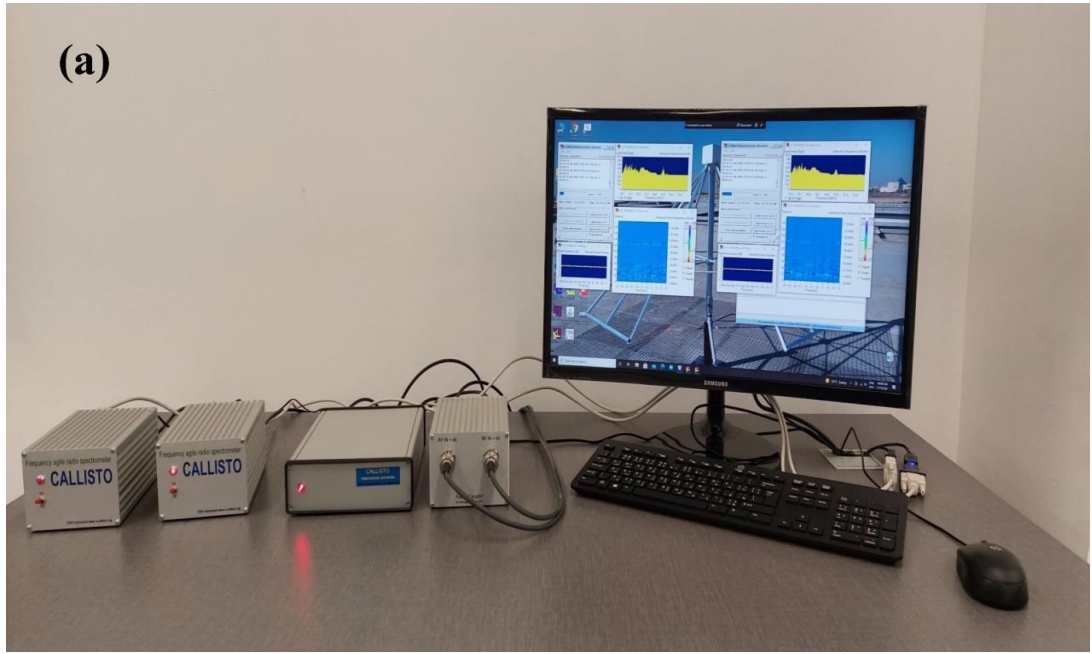

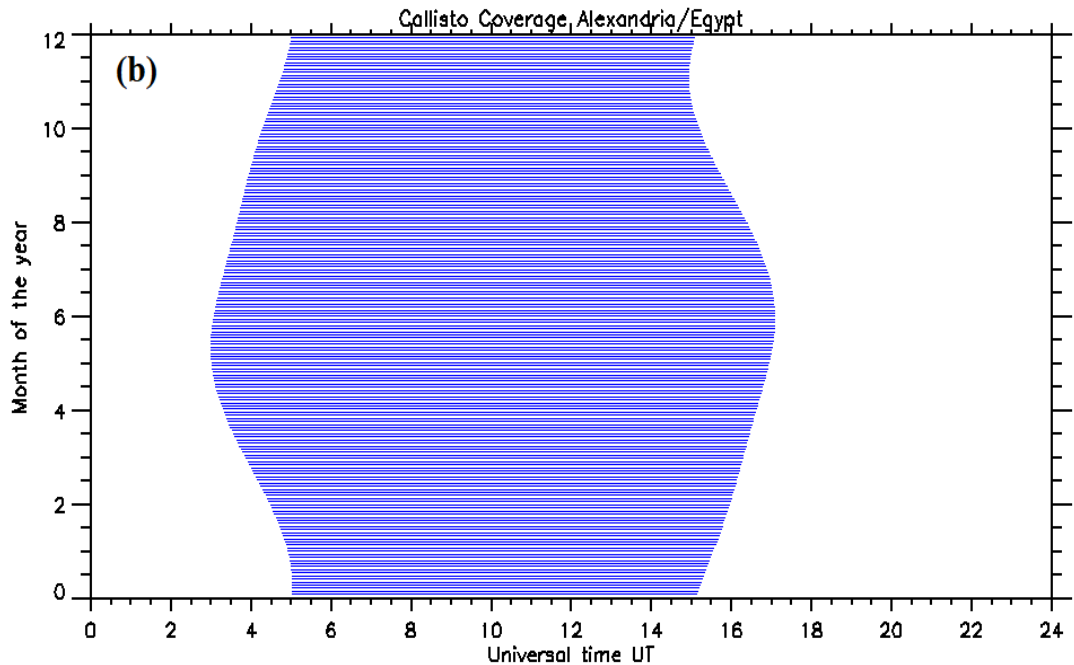

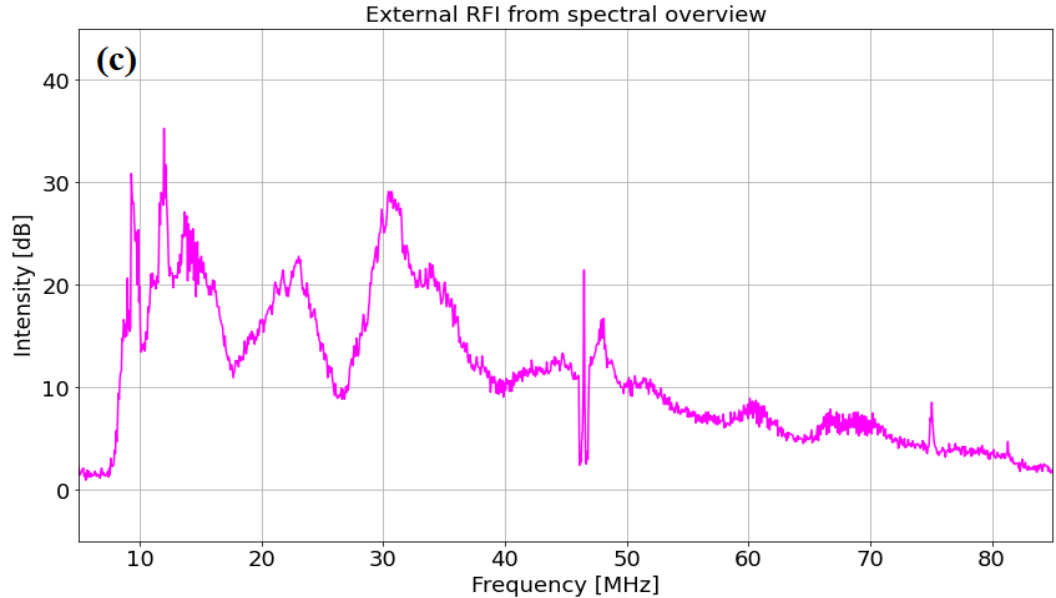

Fig. 2: (a) Fully deployed CALLISTO system in the control room of Space Environment Research Lab. The figure contains two CALLISTO agile frequency receivers, a dual-channel heterodynes up-converter, a power coupler, a standard computer with a windows 10 operation system, and other peripherals. (b) depicts the CALLISTO station's observational sweep at EJUST, Egypt- Alexandria. (c) represents a plot of the external RFI obtained from the raw spectra data. Strong RFI was realized within 10- 20 MHz and 40- 50 MHz.

The left and right polarized receivers are designed to operate in a frequency range of 10 – 80 MHz with a time resolution of 0.25 sec at 200 channels per spectrum (800 pixels/sec). The size of each spectrum is a 200 × 3600 matrix with the rows representing frequency and the columns as time. The system employs its software, supplemented by third-party tools, namely, i) the Frequency-generator, which prepares the essential files for system configuration and minimizes the stress in viewing spectral files. It is also used for performing spectral scan/ overview to obtain the desired frequency range; ii) System Scheduler is used to set the sunrise and sunset times; iii) WWWGenie tool plots the output of the receivers by utilizing the output files from the light curves. The light curve represents the instantaneous output from the receiver in each frequency; iv) Java FITS Viewer for viewing the spectrum files and v) FTP-Watch Dog, which is responsible for uploading the FITS files to the e-CALLISTO central server at FHNW, Switzerland. The CALLISTO software, which controls all the fundamental operations, generates four different real-time output files (image data as a FITS file, activity log and light- curve as TXT files, and spectral overview as a PRN file). Fig.2 (a) depicts a complete view of the CALLISTO system in the control room. The system comprises two CALLISTO spectrometers, a dual-channel heterodynes up-converter for calibrating the frequency ranges according to the geographical location, a power supply, 90° quadrature hybrid for converting linear polarization into circular polarization, and a standard computer with internet. Fig.2 (b) presents the instrument's time coverage at the station throughout the year, and it is worth noting that the time fluctuates in response to the seasons. Fig.2 (c) shows the plot of the external radio frequency interference (RFI) estimated using the minimum value of the spectrum measurement as a reference to assess the suitability of the system's frequency range. It must be pointed out that the minimum value of sky measurement was used as a reference since absolute calibration with an LWA antenna is presently impossible. Nonetheless, it is evident from the plot that the frequency ranges, 10 -20 MHz and 40- 50 MHz, contain strong RFI of values within 10 to 36 dB. Hence, the frequency range above 50 MHz is expected to be suitable for radio burst measurement.

### 3.0 Preliminary Results

*3.1 Observations, analyses, and case studies.*

Following the installation of the LWA at the Egypt-Japan University of Science and Technology in Alexandria (N030⁰: 51.69, E029⁰: 33.67), the calibration and configuration processes were conducted remotely with the help of Christian Moisten, allowing the instrument to observe radio spectra in the same way as any other radio spectrometer. In carrying out the configuration settings, the geographical location of the antenna was adjusted from the configuration file according to the specification of the LWA. The frequency generator was used to perform a spectral scan/ overview to obtain the system's best frequency range [10- 80 MHz]. The system scheduler and Perl were downloaded and installed. The former triggers the upload of the FITS files into the central server using the latter and the FTP-Watch Dog. The sunrise time [03:00 UT] and the set sunset time [17:00 UT] for the instrument were set using the system scheduler (see Fig.2 b). It is noteworthy that these settings were conducted remotely by Christian Moisten for the left and right polarized frequency receivers after he was granted remote access right to the computer. The radio spectrometer started functioning on August 14, 2021. It was mounted and welcomed into the global network of CALLISTO spectrometers (e-CALLISTO) after the first light of a tiny type III radio burst was observed on August 20, 2021 (http://www.e-callisto.org/StatusReports/status_91_V01.pdf). Although solar activity is very low, the observation justified the capacity and sensitivity of the spectrometer in detecting solar radio bursts. A week after the first observation, type III and type V events were recorded on August 27, 2021. On October 09, 2021, a type II radio burst event was observed. In this vein, the case study will focus on the latter and one selected type III radio burst, specifically the burst recorded on August 27, 2021.

Fig.3 (a) presents a type II radio dynamic spectrum detected by the instrument between 06:30 UT and 07:00 UT on October 09, 2021. It is worthy to note that the radio burst lasted for ~30 minutes. Thus, two separate type II radio bursts were recorded at different frequencies within 30 minutes. However, the spectrum presented in this study was for the first 15 minutes. The type II burst exhibits both fundamental and harmonic emissions. The former and the latter both possess a starting frequency of ~80 MHz and an ending frequency of ~25 MHz. Their respective onset times are 06:35 UT and 06:38 UT. The fundamental band is relatively dominant compared to the harmonic band. A vigorous survey was conducted from several ground stations within the e-CALLISTO network to assess the status of the type II event. It was ascertained that stations not limited to ALMATY-India, MRO-Finland, SRI-Lanka, ASSA-Australia, AUSTRIA-UNIGRAZ, and SWISS-IRSOL also observed the radio burst at the same time.

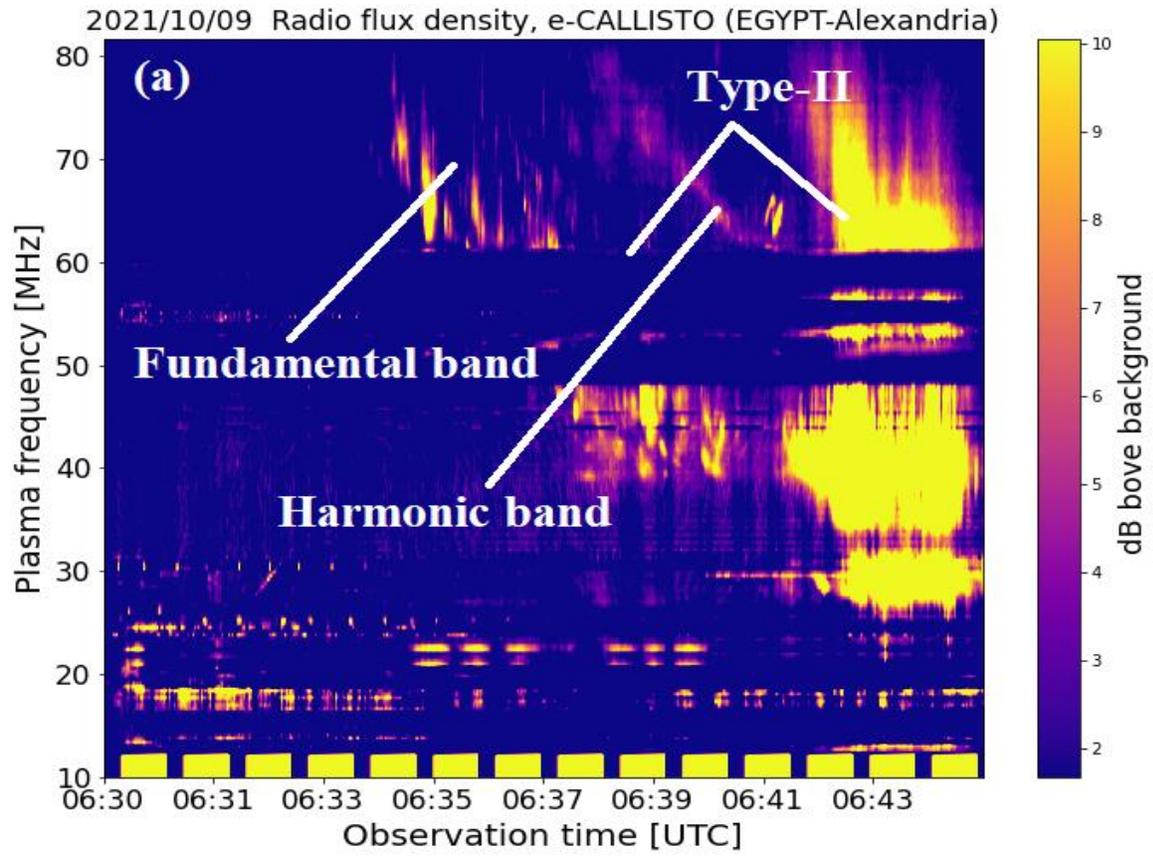

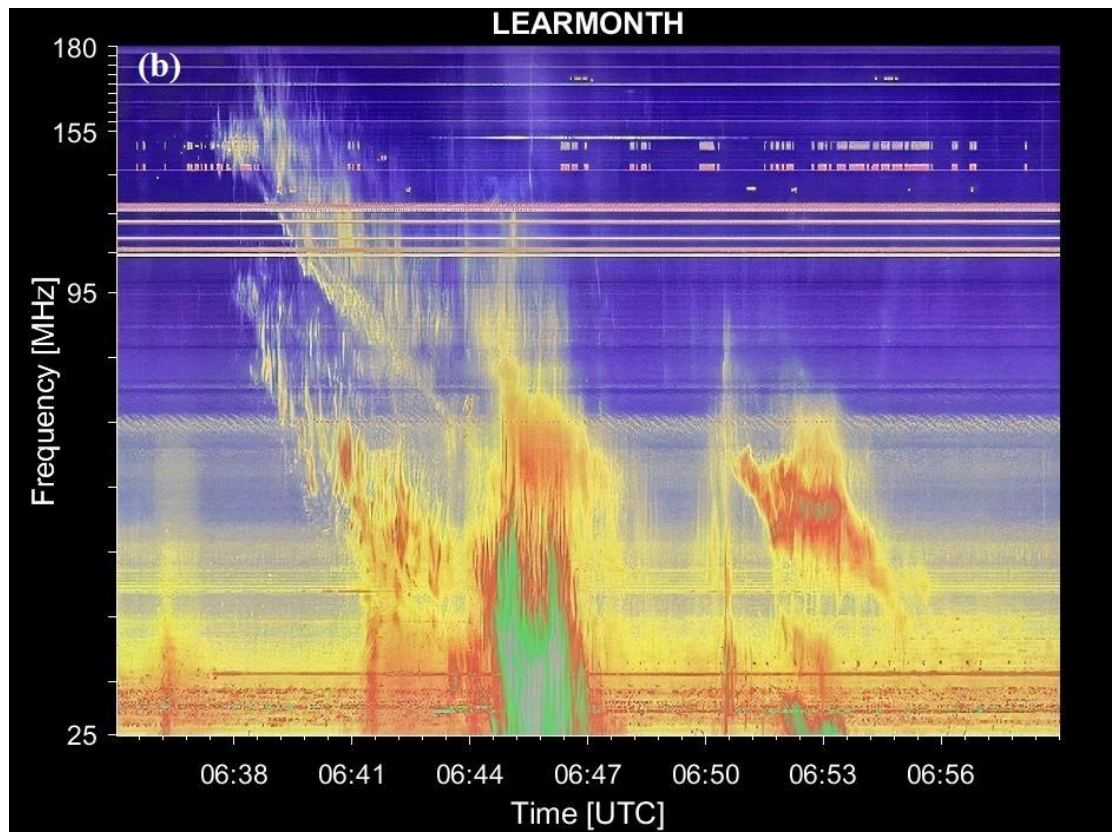

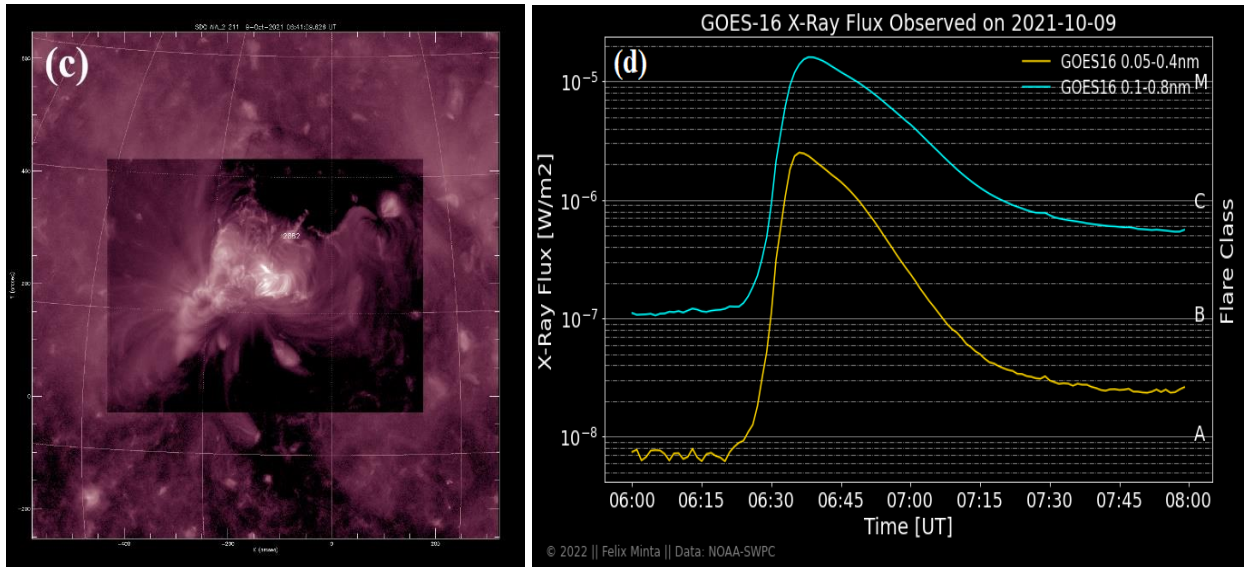

Fig. 3: (a) Type II radio dynamics spectrum observed by the Egypt- Alexandria CALLISTO instrument on October 09, 2021, between 06:30 UT and 06:45 UT. The fundamental and harmonic components have a starting frequency of ~80 MHz at 06:35 UT and 06:38 UT, respectively. (b) shows the spectrum of the type II radio burst recorded by the Learmonth observatory (180- 20 MHz). The dynamic spectrum has similar morphology as the former, with fundamental and harmonic bands present. The starting and ending frequencies of the fundamental band are ~97 MHz and 25 MHz, respectively. The harmonic band began at 180 MHz and ended at 25 MHz. (c) SDO/AIA 211 Å filter observation of solar flare activity at 06:41 UT. The embedded rectangle on the solar disk image shows NOAA-AR 12882 of the solar flare located at N18E08. (d) shows the GOES soft X-ray curves of the solar flare classification. The gold and cyan curves represent the two channels' spectral wavelength bands (0.05nm- 0.4nm, 0.1-0.8nm). The highest peak indicates the M1.6 flare class.

Further, the survey extended to other radio stations outside the terrain of the e-CALLISTO network. It was realized that a more intense and complex spectrum was observed between 06:30 UT and 07:00 UT by Gauribidanur Radio Heliograph (https://www.iiap.res.in/centers/radio#Radio_heliograph). Learmonth solar radio observatory also detected the type II burst under review. Fig.3 (b) shows a similar radio dynamic spectrum sourced from the Learmonth observatory. The starting and ending frequencies of the type II burst observed by the radio telescope are 180 MHz and 25 MHz, respectively. Thus, the fundamental frequency band was observed at ~97 MHz (~06:35 UT) and ended at 25 MHz (~06:45 UT), while the harmonic frequency band commenced at ~180 MHz (~06:33 UT) and elapsed at 25 MHz (~06:50 UT). The Radio Heliograph also observed a group of type III at ~06:28 UT that preceded the type II burst. It is worth mentioning that the fundamental and harmonic components are present in both the two type II bursts. In addition, all the type II bursts are preceded by type III bursts.

Checks from the Atmospheric Imaging Assembly (AIA; Lemen et al., 2012) onboard Solar Dynamic Observatory (SDO; Pesnell et al., 2012) and GOES satellite revealed that the event was associated with a solar flare, occurring between 06:19 UT and 06:53 UT located at N18E08 within NOAA-AR 12882 (see Fig 3 (c and d). The solar flare peak coincides with the M1.6 flare class (cyan curve) at 06:38 UT. Inspecting the LASCO- C2 (STEREO- COR2A) CME catalog hosted by the Solar Eruption Event Detection System (SEEDS) to confirm the existence of a CME within a ~1-hour window; it was noticed that the type II burst was associated with a CME event. The CME first appeared in LASCO- C2 as a disk event at 07:12 UT and was seen in STEREO- COR2A at 07:23 UT as a limb event (see Fig.4 (a and b)). The shock front (leading edge) speeds of the CME in STEREO (LASCO) obtained from the linear fit of the height-time data were ~978 (386) km/s, which are very close to the stated values in the catalog (Fig.4c). From the linear fitting of the height-time measurement shown in Fig.4(d), the CME speed obtained from STEREO and LASCO were ~884 km/s and ~320 km/s, respectively. The shock front and CME speeds are comparatively lower in LASCO- C2 FOV than STEREO. Lower speeds in the former indicate that the CME suffered a projection effect since it appeared as a partial halo/ disk event.

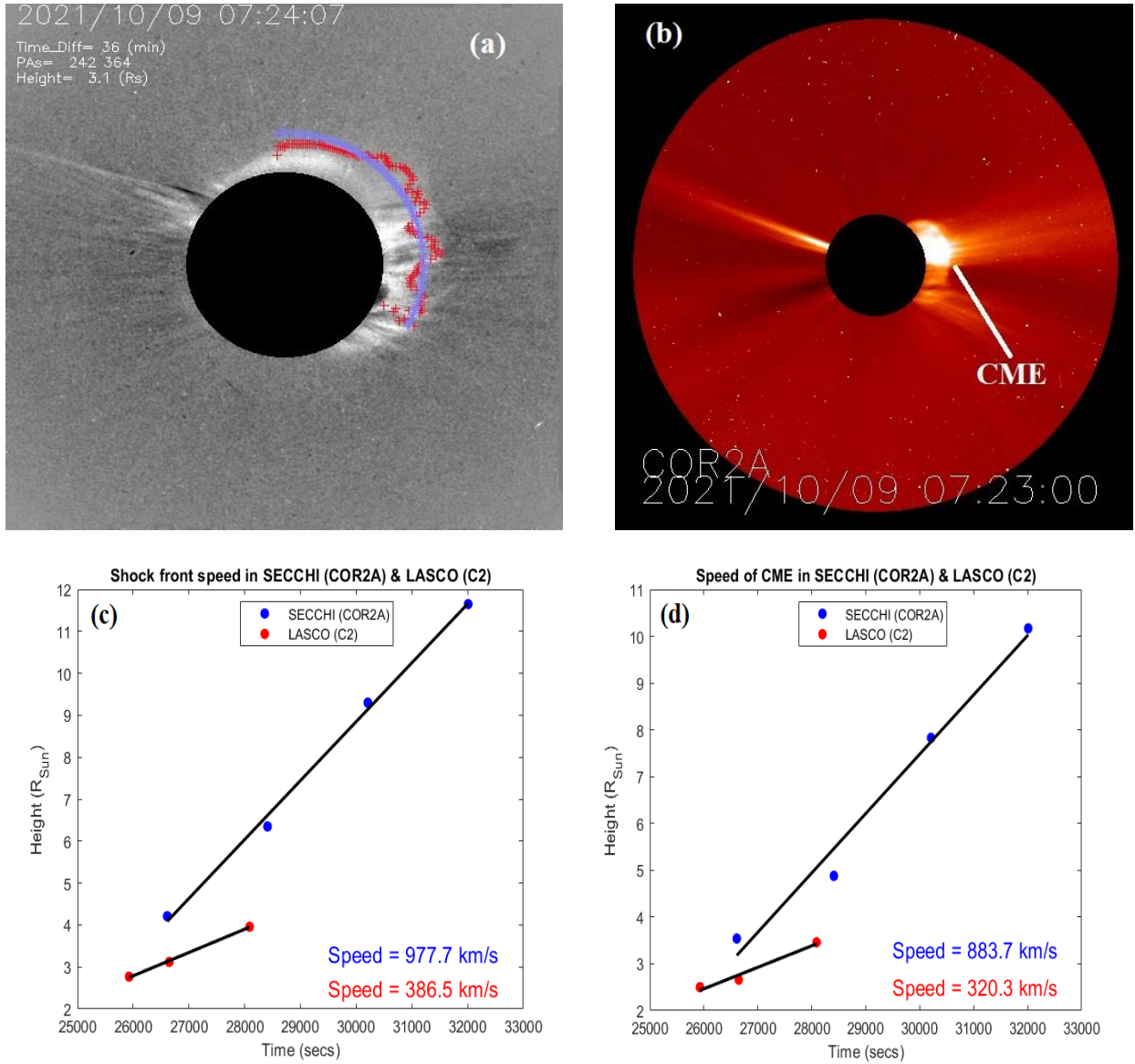

Fig. 4: The running difference image of the CME event as observed by LASCO- C2 (a) and SECCHI- COR2A package onboard STEREO (b) on October 9, 2021. The event first appeared on LASCO at ~7:12 UT and was seen on STEREO- COR2A at ~7:23 UT. The CME was a partial halo event in LASCO and appeared as a limb event in STEREO. The blue curve shows the position of the leading edge, and the red curve indicates the estimated outline of the leading edge. The speed was measured at the shock front (leading edge) in STEREO (blue) and LASCO (red) with the SEEDS data (c). The actual CME speed in STEREO (blue) and LASCO (red) (d).

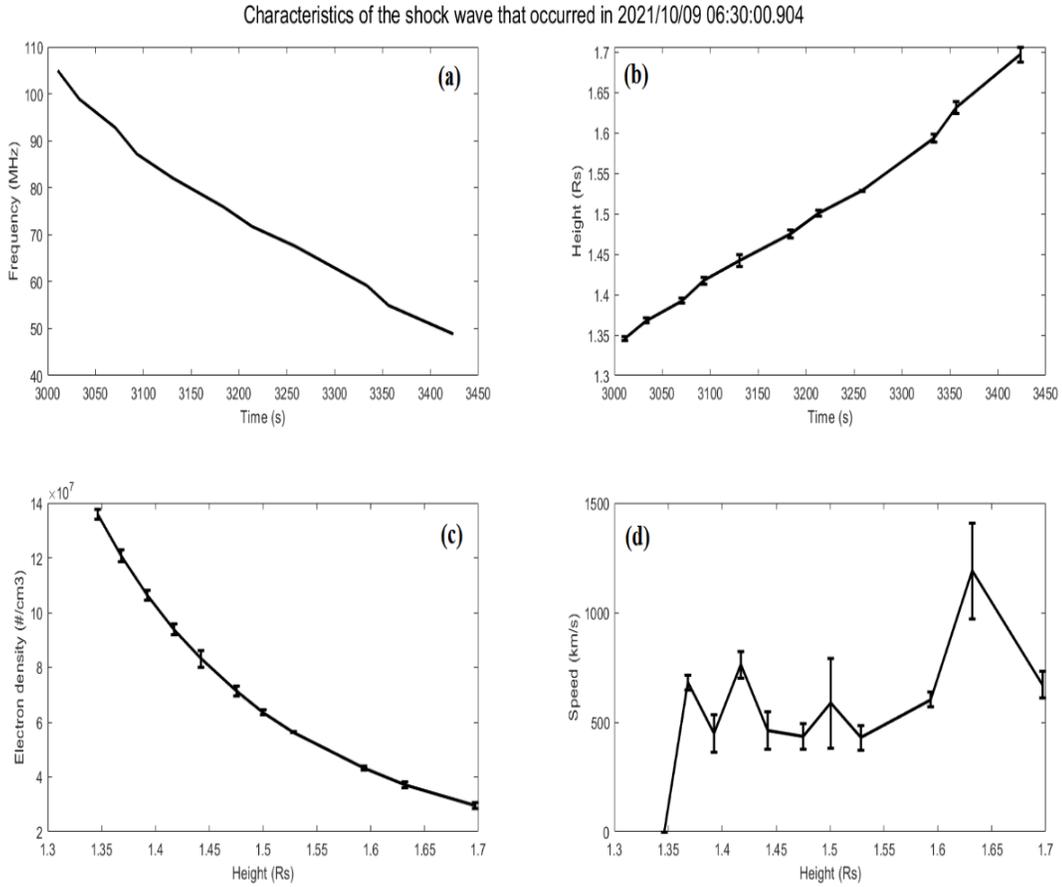

Fig. 5: Shock characteristics estimated from the dynamic spectra. Panel (a) shows a plot of the fundamental frequency against time. Panel (b) shows a plot of the height–time profile. Panel (c) presents the electron density relation with shock height, and panel (d) presents the CME/shock speed variation with heliospheric distance (expressed in solar radii). The standard error was used as a metric for estimating the uncertainties.

The physical conditions were further analyzed with the type II burst by deploying the plasma frequency relation defined as,

$$N_e[cm^{-3}] = (9 \times 10^3)^{-2} f_p^2 [kHz] \qquad (1)$$

to obtain information about the plasma density at which the emission occurs. Where $N_e$ and $f_p$ are the electron density and plasma frequency, respectively.

Consequently, the Newkirk electron density model (Newkirk, 1961) was used to estimate the variation of shock height at which the emission occurs,

$$N_e = \alpha \times N_{eo} \times 10^{4.32(R_\odot/R_s)} \qquad (2)$$

where $\alpha$ is the fold number describing the electron density profile at quiet, i.e., $\alpha = 1$, and active, i.e., $\alpha = 4$ across the equatorial regions, $N_{eo}$ is given as $4.2 \times 10^4$, $R_\odot$ represents the solar radii, and $R_s$ denotes the heliospheric distance. Further, the CME/ shock speed was estimated to assess the type II co-relational association with CME kinematics and dynamics.

The standard error (SE) expressed below was adopted as a metric for assessing the uncertainties in the measurements,

$$\text{SE} = \frac{\sigma}{\sqrt{n_m}} \qquad (3)$$

where σ and $n_m$ represent the standard deviation and number of measurements, respectively.

Fig 5 (a) shows a plot of the fundamental frequency against time. The estimated frequency ranges between 105 MHz and 49 MHz. The mean and median are ~77 MHz and ~75.9 MHz and are consistent with the ranges provided by (Alielden, 2019; Minta et al., 2022). Panel (b) of Fig. 5 shows the height-time profile of the CME extracted from the dynamic spectrum. The estimated shock height ranges between 1.34 $R_s$ and 1.68 $R_s$. The mean and median are 1.49 $R_s$ and 1.47 $R_s$, respectively. The range of shock height obtained in the study conforms with the ranges reported in previous studies (see Gopalswamy et al., 2012; Gopalswamy et al., 2013; Vasanth et al., 2014; Kishore et al., 2016; Kumari et al., 2019; Nedal et al., 2019; Umuhire et al., 2021a; Umuhire et al., 2021b). Panel (c) of Fig. 5 represents the electron density variation across the shock. The estimated electron density across the shock lies between $1.420 \times 10^8$ and $3.2996 \times 10^7$ cm$^{-3}$, respectively. It is worthy to note that this range is consistent with earlier findings on electron densities (see Cunha-Silva et al., 2014; Alielden, 2019; Minta et al., 2022). From Fig.5 (d), the starting CME/ shock speed estimated from the metric type II radio burst was ~605 km/s at a corresponding heliospheric distance of ~1.68 $R_s$. The CME speed obtained from the type II burst is comparable to the CME speeds from the white light measurements shown in Fig.4. This emphasizes that shock formation can occur at heights below 1.5 $R_s$, and type II radio bursts with low starting frequencies can also lead to CME/ shock formation at higher heights (see Gopalswamy et al., 2013; Minta et al., 2022). Thus, the type II burst can complement white light observations in studying CME kinematics. The estimated CME/ shock speed associated with type II radio burst is consistent with the mean CME/ shock speed (610 km/s) reported by Gopalswamy et al., (2008).

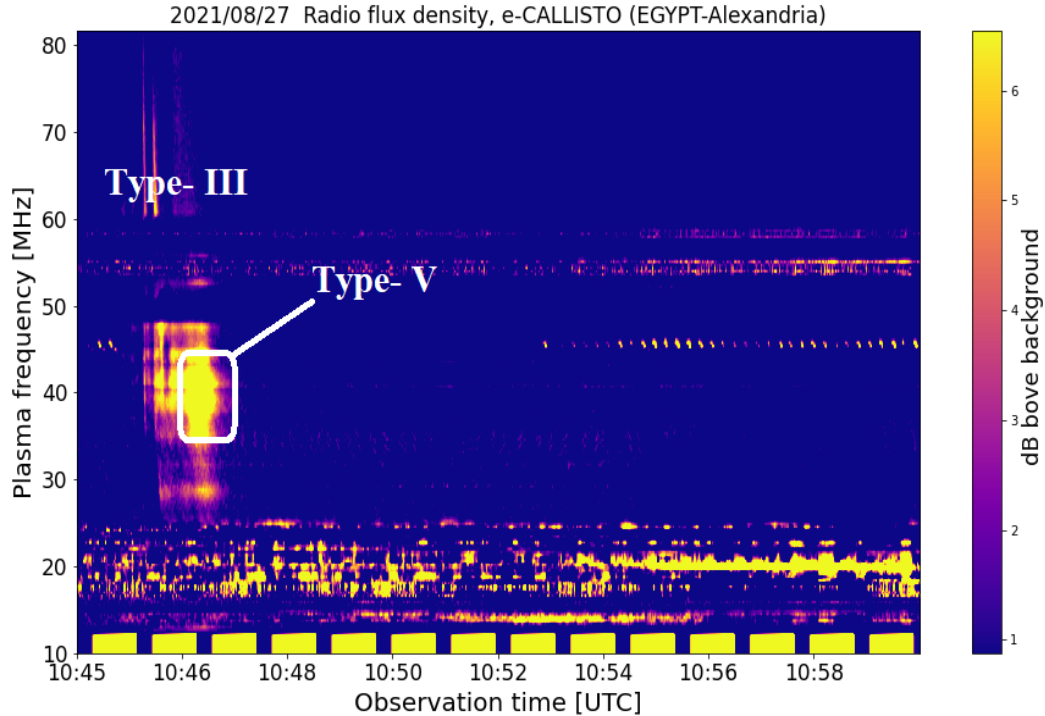

Fig. 6: Type III radio burst observed on August 27, 2021, between 10:45 UT and 11: 00 UT. The starting and ending frequencies are ~80 MHz and ~25 MHz. The type III burst was immediately followed by a type V which lasted for ~ 1 minute.

Fig.6 shows a spectrum of a type III radio burst detected by the instrument on August 27, 2021, from 10:45 UT to 11:00 UT. The upper and lower frequency cut-offs are ~80 MHz and ~25 MHz, respectively. The type III event was condensed with a type V radio burst between 10:46 UT and 10: 47 UT. However, the type V appears to be the dominant feature of the radio spectrum. It was noticed that the same event was detected by CALLISTO- BIR, Brazil, which operates between 20MHz and 100 MHz frequency range (http://soleil.i4ds.ch/solarradio/qkl/2021/08/27/BIR_20210827_104500_01.fit.gz.png). The upper and lower cut-off frequency of the radio burst detected by the station mentioned above was 60 MHz and 10 MHz, respectively. After vigorous checks from various space weather monitoring and forecasting databases, it was ascertained that the event was neither associated with CME nor solar flare.

**4.0 Summary, conclusion, and future objectives.**

The CALLISTO system deployed in Alexandria- Egypt, hosted by the Egypt- Japan University of science and technology, has detected four successive radio bursts, 1 type II, 2 type IIIs, and 1 type V, at different times. The system specification and configuration are described in section 2 of this paper. The complete system consists of a cross dipole LWA located at N030º: 51.69, E029º: 33.67, two frequency receivers, dual-channel heterodyne up-converters, and a standard windows-based computer. The configuration and calibration were done with the help of the principal inventor, Christian Monstein. A detailed case study was conducted on two selected radio bursts (type II and type III) to verify the instrument's capacity among other spectrometers.

During the survey, it was observed that the type II was associated with a CME event which first appeared in LASCO at 07:12 UT and was seen in STEREO at 07:24 UT. The CME in the former was partly halo (position angle of 303º) with a shock front speed of 386 km/s. In the latter, the CME appeared as a limb event with a shock front speed of 978 km/s. The estimated CME speed from the type II was 605 km/s which was significant. Additionally, the type II was associated with M1.6 solar flare within NOAA-AR 12882 at N18E08. The solar flare starts at 06:09 UT, peaks at 06:38 UT, and ends at 06:53 UT. Contrary, the type III and type V bursts observed between 10: 45 and 11:00 UT were not associated with CME and solar flare.

It is worthy to note that other ground stations at different geographical longitudes and latitudes also detected all the spectra recorded by the instrument during the low solar activity. Furthermore, it can be pinpointed that the Egypt- Alexandria and the ALGERIA- CRAAG CALLISTO stations are the only stations operating in the northern part of Africa. CALLISTO stations located in the eastern part of Africa, specifically Kenya and Ethiopia, are not operating due to lack of maintenance. The western part of Africa has no CALLISTO station, while the southern part of Africa has one working station located in South Africa. Thus, the Egypt- Alexandria station serves as a reference station and augments the already operating stations within the e-CALLISTO network to address the challenges of data gaps faced by developing countries, particularly in the African continent. Hence, the CALLISTO system deployed by E-JUST becomes crucial for space weather monitoring and forecasting over the region and other parts of the world. It is anticipated that radio frequency interference (RFI) mitigation mechanisms will be incorporated within the system to curb their interference with the radio spectrographs in the future. Also, it is expected that more radio bursts will be detected during solar maximum, and machine learning algorithms will be implemented to ascertain the association between the physical conditions of radio bursts and other space weather phenomena, including CMEs and solar flares.

**Acknowledgments.**

This work is supported by the Egyptian Academy of Scientific Research and Technology (ASRT), Science UP Capacity Building-Central Labs Project ID: 6379. Many thanks to the e-CALLISTO data center hosted by FHNW, Institute for Data Science-Switzerland and managed by Christian Monstein, the principal inventor of CALLISTO; Learmonth Solar Radio Observatory; NOAA-SWPC; GOES; SOHO/LASCO, STEREO, for their policies on open data accessibility. The first author expresses his profound gratitude to Mohammed Nedal (Institute of Astronomy with National Astronomical Observatory, Bulgarian Academy of Sciences, Sofia, Bulgaria) for his intuitive contribution to conducting this research. Our heartfelt appreciation to the anonymous reviewers for their helpful and encouraging comments in improving this manuscript.